\documentclass[preprint2,11pt]{aastex}

\newcommand\puncspace{\ifmmode\,\else{\ifcat.\C{\if.\C\else\if,\C\else\if?\C\else%
\if:\C\else\if;\C\else\if-\C\else\if)\C\else\if/\C\else\if]\C\else\if'\C%
\else\space\fi\fi\fi\fi\fi\fi\fi\fi\fi\fi}%
\else\if\empty\C\else\if\space\C\else\space\fi\fi\fi}\fi}
\newcommand\SP{\let\\=\empty\futurelet\C\puncspace}

\newcommand{\mdot}{$\dot{M}$\SP}
\newcommand{\lx}{$L_x$\SP}
\newcommand{\freq}{$\nu_{kHz}$\SP}
\newcommand{\mm}{M\'endez }

\slugcomment{Accepted by ApJ, 2 Februrary 2000}

\shorttitle{QPOs and Spectra}
\shortauthors{Ford et al.}

\begin{document}

\title{Simultaneous Measurements of X-Ray Luminosity and Kilohertz
Quasi-Periodic Oscillations in Low-Mass X-Ray Binaries}

\author{Eric C. Ford\altaffilmark{1}, Michiel van der
Klis\altaffilmark{1}, Mariano \mm\altaffilmark{1,2}, Rudy
Wijnands\altaffilmark{1,3}, Jeroen Homan\altaffilmark{1}, Peter G.
Jonker\altaffilmark{1}, Jan van Paradijs\altaffilmark{1,4}}

\email{ecford@astro.uva.nl}

\altaffiltext{1}{Astronomical Institute, ``Anton Pannekoek'',
University of Amsterdam, Kruislaan 403, 1098 SJ Amsterdam,
The Netherlands}
\altaffiltext{2}{Facultad de Ciencias Astron\'omicas y Geof\'{\i}sicas,
Universidad Nacional de La Plata, Paseo del Bosque S/N, 1900 La Plata,
Argentina.}
\altaffiltext{3}{MIT, Center for Space Research, Cambridge, MA 02139}
\altaffiltext{4}{University of Alabama in Huntsville, Department
of Physics, Huntsville, AL 35899}

\begin{abstract}

We measure simultaneously the properties of the energy spectra and the
frequencies of the kilohertz quasi-periodic oscillations (QPOs) in
fifteen low mass X-ray binaries covering a wide range of X-ray
luminosities. In each source the QPO frequencies cover the same range
of approximately $300$ Hz to $1300$ Hz, though the sources differ by
two orders of magnitude in their X-ray luminosities (as measured from
the unabsorbed 2--50 keV flux). So the X-ray luminosity does not
uniquely determine the QPO frequency. This is difficult to understand
since the evidence from individual sources indicates that the
frequency and luminosity are very well correlated at least over short
timescales. Perhaps beaming effects or bolometric corrections change
the observed luminosities, or perhaps part of the energy in mass
accretion is used to power outflows reducing the energy emitted in
X-rays. It is also possible that the parameters of a QPO model are
tuned in such a way that the same range of frequencies appears in all
sources. Different modes of accretion may be involved for example
(disk and radial) or multiple parameters may conspire to yield the
same frequencies.

\end{abstract}


\section{Introduction}

Many low mass X-ray binaries exhibit quasi-periodic oscillations
(QPOs) in their persistent X-ray flux in the kilohertz range as
revealed by the Rossi X-ray Timing Explorer (RXTE). There are
currently 18 such sources with published results.  Generally two
kilohertz QPOs are observed simultaneously from a given system. In all
cases, the QPOs are separated in frequency by about 250 to 350 Hz. The
QPOs vary over a wide range in frequency. In 4U~0614+09, for example,
the higher frequency QPO has been measured at frequencies between
$449\pm20$ Hz and $1329\pm4$ Hz \citep{vanstraaten00}. For reviews and
references see \citet{vanderklis_rev98b} and
http://www.astro.uva.nl/$^{\sim}$ecford/qpos.html.

The low mass X-ray binaries (LMXBs) which exhibit QPOs come in a wide
variety. Most are persistent sources, but some transients are known
with kilohertz QPOs: 4U~1608-52 \citep{berger96,mendez98a}, Aql~X-1
\citep{zhang98a}, and XTE~J2123-058 \citep{homan99,tomsick99}.  The
two traditional classes of LMXBs, Z and atoll sources \citep{hk89},
have very similar QPOs, though the QPOs in Z-sources tend to have
larger widths and smaller rms fractions. The X-ray dipper 4U~1915-05
\citep{boirin00} also has shown kilohertz QPOs.  In all these systems,
the kilohertz QPO frequencies are very similar, even though the
inferred mass accretion rates differ by orders of magnitude
\citep{vdk_rev97a,vdk_rev97b}.

Here we quantify these comparisons by considering the ensemble of
sources. The main tool is a measurement of the X-ray luminosity in
each system simultaneous with a determination of its kilohertz QPO
frequencies. This approach is inspired by the strong correlation of
QPO frequency and count rate in individual sources. This correlation
is very strict on short time scales
\citep[e.g. 4U~1728-34;][]{strohmayer96}, though on longer timescales
of days to weeks in some sources a single correlation no longer holds
\citep[e.g. 4U~0614+09,][4U~1608-52,]{ford97a,mendez99a}. The same
correlations are present if one considers X-ray flux instead of count
rate \citep{ford97b, zhang98a}. The QPO frequencies are clearly
influenced to some extent by the X-ray luminosity.

Correlations of luminosity and kilohertz QPO frequency provide a
rather direct connection to QPO models. In most current models, the
frequency of one of the QPOs is set by the orbital frequency of matter
in the inner disk \citep{mlp98a,lai98,sv99,ot99b}.  Higher QPO
frequencies are the result of faster orbital frequencies which are in
turn coupled to higher mass accretion rates.

In the following we present simultaneous measurements of kilohertz
QPOs and energy spectra in LMXBs. Section~2 details the analysis
procedure and results with special notes on each source. Section~3
discusses these results in context with current QPO models.

\section{Analysis \& Results}

In this analysis we use data from the RXTE Proportional Counter Array
(PCA), \citep{zhang93}. We consider fifteen sources with kilohertz
QPOs, which includes all sources reported to date except
XTE~J2123-058, 4U~1915-05 and GX~349+2. These latter three sources
have relatively few observations with kilohertz QPOs. For timing
analysis, we construct Fourier power spectra from the high-time
resolution modes of the PCA with Nyquist frequencies of typically 4096
Hz. We fit these power spectra for QPO features in roughly the
200--2000 Hz range.  For intervals where a QPO is detected, we perform
spectral fitting using the 16 sec resolution `Standard 2' mode PCA
data.


In the sources where the QPOs are strong (e.g. 4U 1608-52), the QPO
features are significantly detected in a time interval of 64 sec or
less. In these cases we have chosen representative intervals and
performed the spectral fitting on the identical time interval where
the QPOs are detected.  In other sources (e.g. 4U 1705-44) many power
spectra from short time windows must be added before the signal to
noise improves to the level where the QPOs are detected. In such cases
the spectra are well measured on much shorter time scales and we
select an interval (typically 64 sec duration) in the middle of the
interval where the QPOs are detected. There are no large count rate or
color variations within these intervals so this procedure is accurate.
In the case of Z sources, the QPO frequencies have been measured as a
function of $S_z$, the position along a track in the X-ray
color--color diagram or hardness--intensity diagram
\citep[e.g. GX~17+2,][]{wijnands98a}. In these cases we perform
spectral fitting on matching intervals of $S_z$, using the same
observations where the timing analysis was performed.

In spectral fitting we use only the top of the three xenon/methane
layers of the proportional counter units (PCUs) to reduce systematic
effects. We also do not include events in the uppermost
anticoincidence propane layer. We use all of the five PCUs when
available, though in a few cases one or more PCUs were off, and we
performed spectral fitting on the subset of detectors that was on. We
use the background estimation tool pcabackest v2.1b, response matrix
generator pcarsp v2.38 and the standard XSPEC v10.0 fitting
routines. Since the response is not well calibrated at low energy we
ignore standard mode 2 PCA channels 1--3 ($<2.4$ keV for gain epoch
three: 15 April 1996 to 22 March 1999). We also ignore channels above
55 ($>22.4$ keV, gain epoch three) since the background dominates
there even in the brighter sources.  We have ignored the HEXTE data,
since this provides no constraints on the spectral fit for the short
intervals we consider here.

We have chosen to describe the continuum spectra in terms of the
following model components: a power law, a blackbody, and a Gaussian
line at roughly 6.4 keV, all absorbed with an equivalent hydrogen
column density. This model, which is purely phenomenological, is often
used in the literature \citep[e.g.][]{cs97,wsp88} but is not intended
as a physically self-consistent representation of the physical
processes at work. All the parameters of the models are allowed to
float (though in some cases the width of the Gaussian line is
fixed). The reduced $\chi^{2}$ values are close to one in all
cases. There is no evidence for a roll-over at high energies,
indicating that a power law is a sufficient description at least up to
our cutoff energy of $\sim22$ keV.

From the model fits we calculate several parameters, the most
important here being the total flux from 2 to 50 keV. We report the
unabsorbed flux, which is corrected for the effect of absorption at
low energies by the interstellar medium and represents the actual flux
emitted by the source.  We take the unabsorbed 2--50 keV flux as
some indication of the bolometric flux of the source, though it is an
obviously flawed indicator since the spectra are unmeasured below 2
keV and above $\sim22$ keV. Observations with the Beppo-SAX
instruments, however, have good statistics over a much wider energy
range (0.1 to 200 keV). Beppo-SAX observations of 4U 0614+09
\citep{piraino99} and X1724-308 \citep{guainazzi98} indicate that the
model we employ here provides an accurate description of the
spectra. In the 4U~0614+09 observations, \citet{piraino99} find an
accurate fit to the spectrum with a blackbody at $kT=1.45$ keV, a
powerlaw with photon index 2.33 and a line at 0.71 keV that carries
1\% of the total flux, all absorbed by an equivalent neutral hydrogen
column of $3.3\times10^{21}$ cm$^{-2}$. This spectral description is
similar to the one used here.

In reporting here the unabsorbed 2-50 keV flux we tend to
underestimate the actual flux because of the truncation in energy. By
truncating at 2 keV we underestimate the flux that would have been in
the blackbody by roughly 2\% to 20\% in these spectra.  By stopping
the integration at 50 keV we also underestimate the flux at high
energy, which in principle can be a large amount because of the hard
tails in some sources \citep[c.f.][]{barret94}. The observations
considered here, however, did not find any source in an extremely hard
state.  We estimate that we typically loose about 2\% of the flux in
the power law by stopping the integration at 50 keV. In the hardest
spectra (4U~0614+09 at low flux) we miss about 15\% of the flux.
There is likely a break in the power law at high energy (not included
here) which makes the missing flux somewhat less than that.  In the
Beppo-SAX spectrum mentioned above, the flux from 0.1 keV to 2 keV is
25\% of the total flux and that above 50 keV is 7\% of the
total. Finally, the bolometric flux may be larger than our estimate if
the power law extends to very low energies (though this is physically
not so likely) or if different components are present in the extreme
ultraviolet or soft X-ray band.

To calculate a luminosity, \lx, from the total unabsorbed 2--50 keV
flux, we need to know the source distances. The distances we use here
are quoted in Table~\ref{tbl:dist} along with references. Distances
can be determined in a variety of ways \citep[see][for a
description]{vpm94}.  In the sources showing type-I X-ray bursts, the
distance can be determined from radius expansion bursts where the
luminosity is thought to reach the Eddington limit \citep{lewin93}.
In some bursters, no radius expansion bursts have been observed, and
one derives only an upper limit by assuming the flux is less than the
Eddington limit. We use the upper limits as the actual distances (see
Table~\ref{tbl:dist}) , so that the derived \lx are upper limits in
these cases. One source, 4U~1820-30, is in the globular cluster
NGC~6624 and therefore has a relatively well determined distance.  The
distances to the Z-sources, most of which do not show bursts, are more
uncertain. Most of these sources are likely near the galactic center
\citep{penninx89}. A VLBA parallax measurement of Sco X-1 puts it at
$2.8\pm0.3$ kpc \citep{bradshaw99}. A radius expansion burst was
recently observed from Cyg~X-2, yielding a distance of $11.6\pm0.3$
kpc \citep{smale98}, though results from optical lightcurves put it
substantially closer \citep[see][]{ok99}. The Cyg X-2 fluxes we
measure are consistent with the data from the Einstein Observatory
\citep{cs97} and EXOSAT \citep{schulz99}.

The spectral analysis of Sco X-1 requires a special treatment which
deserves note. In this source, detector deadtime effects are important
since its count rate exceeds 25000 c s$^{-1}$ PCU$^{-1}$. We apply a
correction for nonparalyzable deadtime, which amounts to simply
multiplying the effective exposure time by a factor of about 0.7
\citep{zhang95}. We calculate this factor from the measured rates, a
10 $\mu$sec deadtime appropriate for `Good Xe Events', and a 150
$\mu$sec deadtime appropriate for events registered as `Very Large
Events' in the instrument modes used. This deadtime treatment is
approximate and does not take into account for example gain shifts due
to the high count rates. We compare the flux we derive for Sco X-1 to
that from Einstein observations \citep{cs97}. Relative to GX~17+2,
these fluxes are the the same.

Given the distance, $d$, and the flux, $F_x$, we calculate the the
luminosity as $L_x = 4\pi d^2 F_x$. Note that this assumes the
emission is isotropic.  In quoting luminosities we normalize to an
Eddington luminosity of $2.5\times10^{38}$~erg~s$^{-1}$.  Misestimates
of distance, like the misestimates of flux discussed above, contribute
to a spread in \lx among sources. However, the observed range of \lx
covers over two orders of magnitude and this large of a range cannot
be explained by these effects alone.


\begin{deluxetable}{lll}
\tablenum{1}
\tablewidth{40pc}
\tablecaption{Distances}
 
\tablehead{ \colhead{Source} & \colhead{$D$} & \colhead{Ref.} \\
\colhead{} & \colhead{(kpc)} & \colhead{} }

\startdata
~~~ {\em Atoll sources} & & \\
4U 0614+09  & 3.0 $^{\rm{a}}$ &  \citet{brandt92} \\
Aql X-1     & 3.4 &  \citet{thorstensen78}; [1] \\
4U 1702-42  & 6.7 $^{\rm{a}}$ & \citet{oosterbroek91} \\
4U 1608-52  & 3.6 &  \citet{nakamura89, ebisuzaki87} \\
4U 1728-34  & 4.3 &  \citet{foster86} \\
4U 1636-53  & 5.5 &  \citet{vanparadijs86}; [1] \\
4U 1735-44  & 7.1 $^{\rm{a}}$ &  \citet{ehs84} \\
KS 1731-260 & 8.5 &  \citet{sunyaev90} \\
4U 1820-30  & 7.5 &  NGC~6624; \citet{rich93} \\
4U 1705-44  & 11.0 $^{\rm{a}}$ & \citet{ehs84,cs97} \\
~~~ {\em Z sources} & & \\
Cyg X-2     &  11.6 & \citet{smale98} \\
GX 17+2     &  7.5 & \citet{ehs84,cs97} \\
GX 340+0    &  9.5 & [1] \\
GX 5-1      &  7.4 & [1] \\
Sco X-1     &  2.8 & \citet{bradshaw99} \\

\tablecomments{ The sources and their distances used in this
paper. References for the distances are shown.  [1] is \citep{vpw95}.}

\tablenotetext{a}{This is an upper limit based on burst fluxes. We use
it as the distance in calculating \lx.}

\enddata
\label{tbl:dist}
\end{deluxetable}


The results of the simultaneous spectral and timing measurements are
shown in Figure~\ref{fig:freqlx} as a function of \lx.  Both of the
double kilohertz QPOs are shown; circled symbols are used to indicate
the higher frequency QPO.  The lines connect points in time order, or
in the case of Z-sources, in order along the Z track.

In each case we must identify which of the double QPOs is observed. In
some observations only one QPO is detected. As reported in the current
literature, all sources (except Aql~X-1) are known to have two
QPOs. Both QPOs, however, are not always present in a given
observation.  In 4U 1608-52 the lower frequency QPO peak is generally
the stronger and narrower of the two \citep[see][]{mendez98a}
providing the identification. In 4U~0614+09 there is a robust
correlation between the position in the X-ray color diagram and the
frequency which allows us to determine which QPO is present
\citep{vanstraaten00}. Similarly in other sources the relative
properties of the energy spectra or rms values generally allow a firm
identification of the peak.

The correlation of QPO frequency, \freq, with \lx can be parameterized
as $\nu_{kHz} = A L_x^{\alpha}$. Taking the data of the upper
frequency QPO in Figure~\ref{fig:freqlx} for 4U~1735-44 and
4U~1702-42, we find $\alpha=0.2$ and 0.5 respectively. We note however
that these data on the upper frequency QPO come from observations
widely separated in time. Over long timescales the \freq--\lx
correlations shifts around and parallel lines are observed (see
below). This data may therefore include several tracks of the parallel
line correlations. In the data of 4U~1608-52 we can separate out the
parallel lines \citep{mendez99a} and measure $\alpha$ within each
stretch of correlated data. We find values of $\alpha$ between 0.5 and
1.6 with typical errors of 0.2, using the absorbed 2--10 keV flux
instead of \lx. Note that though these correlations are measured over
a relatively small range in flux, this measurement does not mix up
different tracks. 

Of special note in Figure~\ref{fig:freqlx} are the LMXBs which do not
appear because they do not exhibit kilohertz QPOs: the atoll-type
sources GX3+1, GX9+9, GX9+1 and GX13+1. The upper limits to the rms
fractions of QPOs in these sources are 1 to 3\%
\citep{strohmayer98a,wkp98,homan98}. The luminosities of these sources
lie between the Z sources and other atoll sources \citep{cs97} and
they are an important intermediate class of sources in some models
\citep[see][]{mlp98a}.

We note that only observations in which the LMXBs exhibit QPOs are
reported here. The total range of \lx that a source covers is
generally larger than that in Figure~\ref{fig:freqlx} since kilohertz
QPOs are present preferentially at intermediate fluxes
\citep{mendez99a,mendez99c}. The only known exceptions to this so far
are 4U~0614+09 \citep{vanstraaten00} and 4U~1728-34 \citep{disalvo99}.

\begin{figure*}
\epsscale{1.0}
\plotone{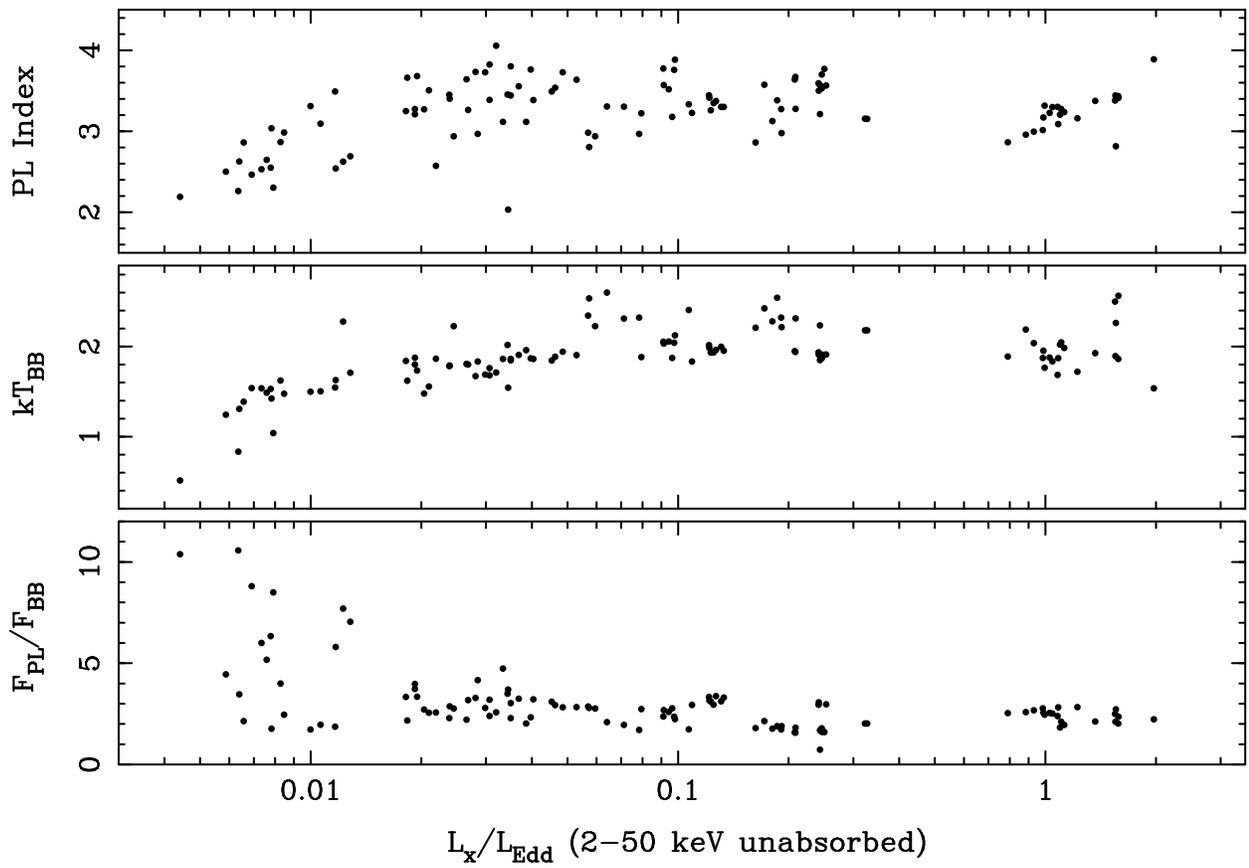} 
\caption{Spectral parameters vs. luminosity. Luminosity is calculated
as in Figure~1. The panels show the index of the power law component
({\em top}), the temperature of the blackbody component ({\em middle})
and the ratio of the absorbed 2--20 keV flux in these two components
({\em bottom}).}
\label{fig:paramlx}
\end{figure*}

Some selected parameters from the spectral fitting are shown in
Figure~\ref{fig:paramlx}. These parameters are similar to those
previously measured for such sources and show that the Z-sources can
be fit by roughly the same spectral model as the atoll sources
\citep[see][]{schulz99,cs97,wsp88}. The ratio of powerlaw to blackbody
flux is 2 to 3 in most cases, i.e. the blackbody is roughly 25\% to
35\% of the total flux \citep[c.f.][]{wsp88}. There is an overall
trend towards harder spectra at lower luminosities, reflected in our
fits. This same trend is seen in previous studies of atoll sources
\citep[e.g.][]{vv94,bg95} and occurs in the emission even up to 100
keV \citep{ford96}. It is also manifest in the patterns in X-ray color
diagrams. The softening at higher fluxes is often attributed to the
effects of thermal Comptonization.

\section{Discussion}

\begin{figure*}
\begin{center}
\epsscale{1.0}
\plotone{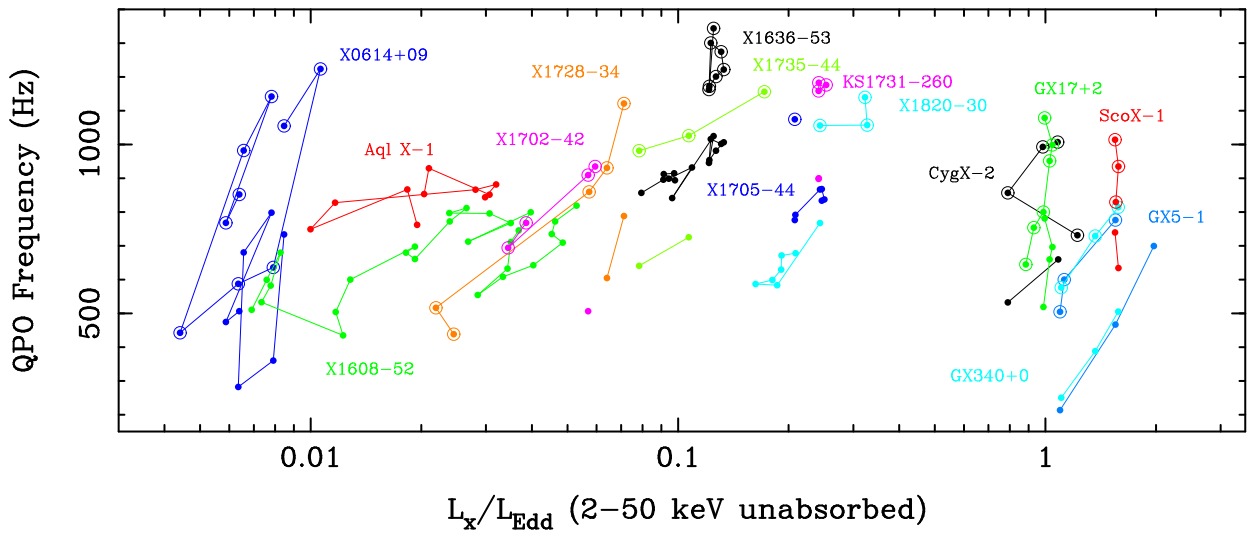}  
\caption{QPO frequency vs. luminosity, $L_x$, in the 2--50 keV
band. $L_x$ is calculated from the distance (Table 1) and the flux in
the model fit for each observation corrected for absorption and
normalized to an Eddington luminosity of $2.5\times10^{38}$ erg
s$^{-1}$. Circled bullets are QPOs identified as the higher frequency
of the two QPOs; uncircled bullets are the lower frequency QPO.  
~~~~~~~~~ ~~~~~ {\bf NOTE: this figure is in color.}}
\label{fig:freqlx}
\end{center}
\end{figure*}

Within a given low-mass X-ray binary the frequency of the kilohertz
QPOs, \freq, is well correlated with the X-ray flux \citep{ford97b,
zhang98a} or count rate
\citep{strohmayer96,wijnands98d,mendez99a,mendez99b,mendez99c}, at
least on the timescale of about a day. Considering all the binaries as
a group, however, such a correlation does not hold. This is a very
clear feature of Figure~\ref{fig:freqlx}, where \freq covers roughly
the same range of frequencies for sources of widely different X-ray
luminosities, \lx.  All sources have maximum frequencies at roughly
1000 to 1300 Hz, a fact that \citet{zss97} have used to argue that the
maximum \freq is set by the orbital frequency at the marginally stable
orbit.  In addition to the similar maximum \freq, all the sources have
roughly the same minimum \freq and slope of their \freq--\lx
relation. This is the central mystery presented here. How is it that
\lx and \freq are decoupled in the ensemble of systems?

This decoupling has an apparent analog within individual sources.  In
a given system, \freq and \lx (or flux, or count rate) are uniquely
correlated within single observations spanning less than roughly a
day. Between observations more widely separated in time, however, the
correlation shifts and parallel lines appear in the \freq vs \lx
diagram similar to those in Figure~\ref{fig:freqlx}. Note, though,
that these parallel lines in individual sources covers a much narrower
range; flux shifts are a factor of a few at most in individual
sources. 4U~0614+09 was first seen to have such parallel lines
\citep{ford97a,ford97b}, and the same effect is observed in Aql~X-1
\citep{zhang98a}, 4U~1608-52 \citep{mendez99a}, 4U 1728-34
\citep{mendez99b}, and 4U~1636-53 \citep{mendez99c}. There is a
similar effect in Z-sources, where \freq is correlated to the position
on the instantaneous Z-track in the X-ray color diagram
\citep[e.g.][]{wijnands98d, jonker00} while the tracks themselves
shift around in intensity.

One possible solution to the mystery of decoupled \lx and \freq is
that the parameters of the mechanism producing the QPOs are tuned in
such a way that \freq is the same in all systems. As an example
consider the magnetospheric beat-frequency model. A simple version of
the theory predicts that the QPO frequency is set by $\dot{M}/B^2$,
where \mdot is the mass accretion rate and $B$ is the surface magnetic
field strength \citep{as85}. The frequencies could then be the same if
$B$ scaled in such a way that $\dot{M}/B^2$ is constant in all systems
\citep{wz97}. Such a connection between \mdot and $B$ was suggested
previously on other grounds \citep{hk89,pl97}. Other parameters, such
as the neutron star spin, mass or temperature, might be involved as
well, though it is not clear how these would fit into a detailed
model.


The observational data do suggest that \mdot has a role in setting the
QPO frequency. The correlations of \freq and \lx suggest this, in as
much as \lx and \mdot are related (see below). The timing properties
point to a similar conclusion as well. The Fourier power spectra often
show a noise component, whose power decreases with frequency above a
break frequency of roughly 10 Hz.  The break frequency is strongly
correlated with $\nu_{kHz}$
\citep{fk98,vanstraaten00,reig00,disalvo00}. The fact that the break
frequency is thought to be a good indicator of $\dot{M}$
\citep{vanderklis94}, suggests that the frequency of the kilohertz QPO
is also correlated with \mdot. Another timing signal is the QPO at
10--50 Hz \citep[e.g.][]{vanderklis96,fk98,pbk99} which also
correlates with \freq. Thus there are several timing features, all
correlated with one another \citep[see also][]{wk99,pbk99}. In
addition \freq also depends strongly on the energy spectra, sometimes
parameterized as the distance along a track in the X-ray color diagram
\citep[e.g.][]{
vanderklis96,wijnands98a,zhang98c,mendez99a,mendez99c,kaaret99b}. The
implication is that a single parameter underlies these correlations,
and that parameter is likely \mdot.

If there is a connection between \freq and \mdot, one might also
expect a correlation of \freq and \lx, since \lx is some measure of
\mdot. Why then is the range of \freq similar for very different \lx
in Figure~\ref{fig:freqlx}? In the following we consider one logical
possibility: that \lx and \mdot do {\em not} track one another.

Perhaps \lx is simply not a good indicator of the bolometric
luminosity and in fact the bolometric luminosity is similar in all
systems.  In principle \lx could misrepresent the bolometric
luminosity just due to the limited 2--25 keV energy range of the
RXTE/PCA. It is unlikely however that this is a large effect, since
Beppo-SAX measurements from 0.1--200 keV indicate that not much energy
is radiated outside the PCA band for these sources and our spectral
models are applicable \citep{piraino99}. Of course there could also be
strong emission in the unobserved extreme ultraviolet band.

If the emission is not isotropic, the measured \lx will also be an
inaccurate indicator of the total emission. Inclination effects are
one possibility: the lower \lx sources may be viewed at a higher (more
edge-on) inclination making \lx smaller. This effect is well known in
the dipping X-ray systems where the inclination is extremely edge-on
and \lx is low \citep{parmar86}. An added attraction of this scenario
is that it may explain the fact that Z-sources are strong radio
emitters while the atoll-sources are not \citep{fender00}. In this
scenario, the less inclined, higher \lx, Z-sources show strong radio
emission because the radio jet is beamed into the line of sight, while
atoll-sources at higher inclination and lower \lx, are usually not
detected in the radio because the radio jet is more in the plane of
the sky. This may not be the full story, however, since the beaming
would have to be narrow and a search for effects of inclination in the
X-ray spectra with EXOSAT uncovered no evidence that inclination is
important \citep{wsp88}.

A general problem with preserving the same \mdot in all the systems
while changing the observed \lx through anisotropy or bolometric
corrections is that, if all the sources had the same \mdot, they
should all show the same X-ray burst properties. They do not; the
Z-sources, for example, hardly burst at all \citep{lewin93}. In the
low-\lx sources, \mdot is also likely low because the persistent
emission is at least 10 times weaker than in the bursts, some of which
are at the Eddington limit. Assuming the anisotropy is about the same
in the burst and persistent emission, \mdot in these sources is then
likely lower than in the sources near the Eddington limit, such as the
Z sources.

Outflows are another way to decouple \lx and \mdot, and are a
well-known feature of X-ray binaries, as seen for example in the
collimated radio jets \citep{hjellming95,fender99b}. One might expect
that the outflows in the low-\lx systems are stronger than those in
the high-\lx systems to preserve a similar accreted rate in the
various systems. Radio observations, however, suggest that the
opposite is true; the atoll sources are less luminous in radio than
the Z-sources \citep{fender00}.

Another alternative is that part of the \mdot may be ineffective in
determining \freq while not being lost from the system. This could
happen if the mass accretion rate occurs in a two component flow,
radially and through a disk \citep[e.g.][]{gl79,flm89,wijnands96}. The
accretion rate through the disk is primarily responsible for setting
\freq, while the radial flow does not affect \freq but does change
$L_x$ \citep[see][]{kaaret98}. \citet{mlp98a} suggest that the disk
accretion rate is similar in all sources. Matter is `scooped off' into
a radial flow at the magnetospheric radius, and this process is more
efficient in the higher \lx sources because the fields are
stronger. Under this scenario, the QPOs at higher \lx should have a
much smaller rms fraction due to the addition of unmodulated
flux. This represents a problem for this scenario since the rms
fraction apparently does not decrease enough with $L_x$
\citep{ford00}.

All of the above effects may act to decouple \lx and \mdot. As
outlined above, though, no single effect can account for the
decoupling and each has problems.  If \lx and \mdot are unrelated,
\mdot can set the frequency of the QPOs while \lx assumes any value,
as observed.

\acknowledgments

This work was supported by NWO Spinoza grant 08-0 to E.P.J.van den
Heuvel, by the Netherlands Organization for Scientific Research (NWO)
under contract number 614-51-002, and by the Netherlands
Researchschool for Astronomy (NOVA). This research has made use of
data obtained through the High Energy Astrophysics Science Archive
Research Center Online Service, provided by the NASA/Goddard Space
Flight Center.


\clearpage
\onecolumn

\clearpage

\end{document}